%% file: main.tex
\documentclass[lettersize,journal]{IEEEtran}
\IEEEoverridecommandlockouts
\usepackage[utf8]{inputenc} % or \usepackage[latin1]{inputenc}
\usepackage{cite}
\usepackage{amsmath,amssymb,amsfonts}
\usepackage{graphicx}
\usepackage{textcomp}
\usepackage{multirow}
\usepackage{tabularx}
\usepackage{dblfloatfix}
\usepackage{color,soul}
\usepackage{subcaption}
\usepackage{comment}
\usepackage{csquotes}
\usepackage{multicol}
\usepackage[table,xcdraw]{xcolor}
\usepackage{booktabs}
\usepackage{orcidlink}
\usepackage{gensymb}
\usepackage{forest}
\usepackage{tikz-qtree}
\usepackage{cleveref}
\usepackage{tikz}
\usetikzlibrary{shapes,arrows}
\usepackage{commath}
\usepackage[labelformat=simple]{caption}

\title{\huge{
HeatSense: Intelligent Thermal Anomaly Detection for Securing NoC-Enabled MPSoCs}}

\begin{document}

\author{\IEEEauthorblockN{Mahdi Hasanzadeh\orcidlink{0000-0003-3093-1319}\IEEEauthorrefmark{1},
Kasem Khalil\orcidlink{0000-0002-9659-8566}\IEEEauthorrefmark{2},
Cynthia Sturton\orcidlink{0000-0003-3930-7440}\IEEEauthorrefmark{3}, 
Ahmad Patooghy\orcidlink{0000-0003-2647-2797}\IEEEauthorrefmark{1}, ~\IEEEmembership{Member,~IEEE}}

\IEEEauthorblockA{\IEEEauthorrefmark{1}Department of Computer Systems Technology, North Carolina A\&T State University, Greensboro, NC, USA}
\IEEEauthorblockA{\IEEEauthorrefmark{2}Electrical and Computer Engineering Department, University of Mississippi, Oxford, MS, USA}

\IEEEauthorblockA{\IEEEauthorrefmark{3}Department of Computer Science, University of North Carolina at Chapel Hill, Chapel Hill, NC, USA}
 }

\maketitle

\begin{abstract}
Multi-Processor System-on-Chips (MPSoCs) are highly vulnerable to thermal attacks that manipulate dynamic thermal management systems. To counter this, we propose an adaptive real-time monitoring mechanism that detects abnormal thermal patterns in chip tiles. Our design space exploration helped identify key thermal features for an efficient anomaly detection module to be implemented at routers of network-enabled MPSoCs. To minimize hardware overhead, we employ \textit{weighted moving average (WMA) calculations} and \textit{bit-shift operations}, ensuring a lightweight yet effective implementation. By defining a spectrum of abnormal behaviors, our system successfully detects and mitigates malicious temperature fluctuations, reducing severe cases from \textbf{3.00°C to 1.9°C}. The anomaly detection module achieves up to 82\% of accuracy in detecting thermal attacks, which is only 10-15\% less than top performing machine learning (ML) models like Random Forest. However, our approach reduces \textbf{hardware usage by up to 75\% for logic resources and 100\% for specialized resources}, making it significantly more efficient than ML-based solutions. This method provides a practical, \textbf{low-cost solution} for resource-constrained environments, ensuring resilience against thermal attacks while maintaining system performance. 
\end{abstract}


\input{1-Intro}

\input{2-Related_Works}
\input{3-Methodology}

\input{4-Proposed}
\input{5-Results}
\input{6-Conclusions}

\bibliographystyle{IEEEtran}
\bibliography{Paper}
\end{document}

%% file: 1-Intro.tex
\section{Introduction}
\label{sec:Intro}
\IEEEPARstart{I}{n} contemporary computing, Multi-Processor System-on-Chips (MPSoCs) and Network-on-Chips (NoCs) serve as the backbone of modern devices including but not limited to smartphones, health-monitoring systems, and edge systems~\cite{Yan2018ModelingOT}. Although these systems have revolutionized our capacity to process and share information, they also expose critical vulnerabilities. Modern MPSoCs are vulnerable to Hardware Trojans (HTs) that can be inserted at various design stages in order to alter the system's functionality or violate users privacy ~\cite{Cheng2005HeterogeneousMS,RAMJAM2024,Trop2024}. The growing dependency on these systems underscores the urgent need for robust measures to safeguard their integrity and reliability~\cite{Yan2018ModelingOT,HardwareTrojanDetection2023}.

A fundamental component of modern MPSoCs is Dynamic Thermal Management (DTM), which acts as a safeguard to ensure devices operate within safe thermal limits~\cite{huang2024}. By monitoring temperature sensors and dynamically adjusting system performance, DTM protects MPSoCs from overheating while maintaining functionality. However, when the sensors DTM relies on are compromised, this delicate equilibrium can be disrupted, leading to potentially catastrophic outcomes~\cite{huang2024,RAMJAM2024,MitigationHardwareTrojan2024}.

Malicious insertion of HTs—covert circuitries designed to tamper with sensors' thermal readings—can deceive the system into believing it is operating within safe conditions when, in reality, it may be overheating~\cite{Vashist2019SecuringAW,ModelingAnalysisConfluence}. The consequences of such deception are severe: processors might overexert themselves, risking irreversible damage, or systems could unnecessarily throttle performance, resulting in reduced efficiency~\cite{d2021malicious}. Research highlights the severity of this threat, showing that such attacks can degrade system performance by as much as 73\%~\cite{elahi2024}. These vulnerabilities emphasize the need for targeted solutions to detect and mitigate such manipulations~\cite{HardwareTrojanDetectionMitigation2022}. The challenge in addressing HT attacks lies in their subtlety; HT manipulations closely mimic natural temperature variations, making it difficult for traditional security mechanisms to differentiate between legitimate and malicious behavior~\cite{Tiwari2019, MitigationDenialService2016,elahi2024}. 

To address this gap in detection capabilities, we propose an adaptive real-time monitoring mechanism  to identify subtle irregularities in thermal patterns, similar to a vigilant observer spotting a disguised threat~\cite{huang2024}. By integrating the mechanism as a hardware module into NoC routers (the communication backbone of MPSoCs), we establish checkpoints to intercept Trojan-induced anomalies at critical junctures of the system~\cite{Cheng2005HeterogeneousMS} through establishing a balance between precision and efficiency. %Based on a meticulous feature selection,
We identified critical indicators of HT activity, ensuring high accuracy while minimizing computational overhead. Furthermore, we employed hardware-level approximation techniques to maintain agility in real-time applications~\cite{Vashist2019SecuringAW}, enabling lightweight and fast computations. This approach not only enhances security, but also preserves system performance, making it a practical solution for MPSoCs and NoCs.
\begin{figure*}[t]
  \centering
  %\vspace{-10}
  \includegraphics[width=\textwidth]{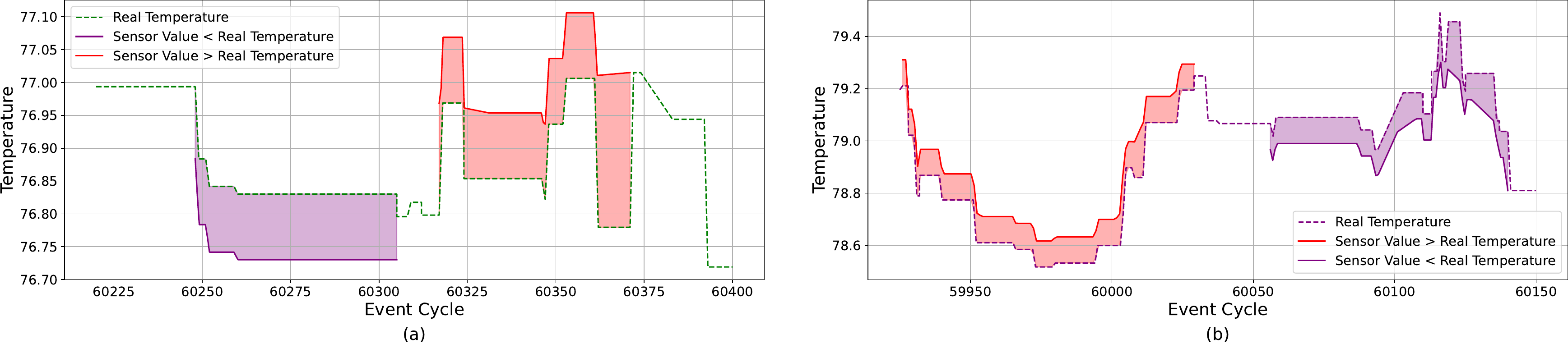}
  \caption{Temperature profiles of router attacks. (a) Attack behavior-1: initial lower captured temperatures followed by increased temperatures. (b) Attack behavior-2: initial higher captured temperatures followed by decreased temperatures. Green dashed line: real temperature; colored solid lines: captured temperatures during attack phases.}
  \label{fig:threat_model}
\end{figure*} 
The contributions of this research are summarized as follows:
\begin{itemize}
    \item Developing a simulation platform to capture thermal behavior of MPSoCs under attack scenarios. 
    
    \item Exploration and selection of key features for accurate and efficient thermal Trojan detection.
    
    \item Development of an anomaly detection mechanism capable of monitoring routers in real-time to detect thermal Trojan activity.
    
    \item Implementation of approximation techniques to reduce computational overhead without compromising detection accuracy.
    \end{itemize}

The rest of this paper is structured as follows. Section \ref{sec:threat} introduces the assumed attack models for MPSoCs and NoCs, emphasizing hardware Trojans targeting thermal sensors. Section \ref{sec:method} details the research methodology, including the use of the Noxim simulator and feature analysis. Section \ref{sec:Proposed} outlines the proposed anomaly detection module, highlighting its real-time monitoring, multi-level classification, and approximation techniques. Section \ref{sec:results} presents experimental results, comparing feature sets and machine learning models. Finally, Section \ref{sec:Conclusions} concludes the paper.

%% file: 2-Related_Works.tex
\begin{table*}[ht]
\centering
\caption{Features extracted and engineered from the collected router-level and network-level data, showcasing metrics relevant to thermal behavior, power utilization, network traffic, and anomaly detection.}
\label{tab:feature_set}
\resizebox{\textwidth}{!}{ % Scales the table to textwidth
\begin{tabular}{|>{\columncolor[HTML]{EAF2F8}}l|>{\columncolor[HTML]{EAF2F8}}p{2.5cm}|c|l|l|}
\hline
\rowcolor[HTML]{D3D3D3}
\textbf{Feature \#} & \textbf{Feature} & \textbf{Value} & \textbf{Feature Description} & \textbf{Category} \\ \hline
\textbf{F1}         & Event Cycle      & 0-5,000,000 (Cycle) & Sequence of steps involved in processing and routing a data packet through the network & Temporal\\ \hline
\rowcolor[HTML]{F5F5F5}
\textbf{F2}         & Core Power       & 0-3 (Watt)          & The power consumption of a processor's core & Core\\ \hline
\textbf{F3}         & Core Temperature & 0-99 (°C)           & The temperature of a processor's core & Core\\ \hline
\rowcolor[HTML]{F5F5F5}
\textbf{F4}         & Core Utilization & 0-100 (\%)          & Processor's computational capacity being used at a given time & Core\\ \hline
\textbf{F5}         & Core Frequency   & 0-3,600 (MHz)       & The operating speed of a processor core & Core\\ \hline
\rowcolor[HTML]{F5F5F5}
\textbf{F6}         & Packet Source    & 0-256               & The origin point of a data packet in a network & Network\\ \hline
\textbf{F7}         & Packet Destination & 0-256             & The intended endpoint for a data packet in a network & Network\\ \hline
\rowcolor[HTML]{F5F5F5}
\textbf{F8}         & Current Router   & 0-256               & A router that logs and records events related to packet transmission & Network\\ \hline
\textbf{F9}         & Flit Type        & \{Header, Body, Tail\} & Specifies the type of flit in network-on-chip communication & Network\\ \hline
\rowcolor[HTML]{F5F5F5}
\textbf{F10}        & Flit Hop Count   & 1-14                & The number of intermediate nodes a flit has passed through & Network\\ \hline
\textbf{F11}        & Flit Sequence No & 0-8                 & A number assigned to each flit to maintain order within a packet & Network\\ \hline
\rowcolor[HTML]{F5F5F5}
\textbf{F12}        & Packet Sequence No & Depends on Packet Injection Rate & A unique identifier assigned to each packet & Network\\ \hline
\textbf{F13}        & Receiving Port   & 0-6                 & The specific port on a router that receives incoming data packets & Router\\ \hline
\rowcolor[HTML]{F5F5F5}
\textbf{F14}        & Departing Port   & 0-6                 & The port from which a data packet leaves a router & Router\\ \hline
\textbf{F15}        & Router's Congestion & 0-100 (\%)        & Measure of the traffic load or congestion level at a specific router & Router\\ \hline
\rowcolor[HTML]{F5F5F5}
\textbf{F16}        & Router's Temperature & 0-90 (°C)        & The current operating temperature of a router & Router \\ \hline
\textbf{F17}        & Router's Temperature 2-Cycle Average & 0-90 (°C) & A smoothed temperature reading calculated by averaging over two cycles & Router\\ \hline
\rowcolor[HTML]{F5F5F5}
\textbf{F18}        & Router's Temperature Running Average & 0-90 (°C) & A continuously updated average of temperature over recent measurements & Router\\ \hline
\textbf{F19}        & Attack Label     & \{0,1,2\}           & A classification of network events identified as potential threats & Security\\ \hline
\end{tabular}%
}
\end{table*}

\section{Attack Model}
\label{sec:threat}
This section details two hardware Trojan attack behaviors targeting thermal sensors in NoC-enabled MPSoCs, based on the model proposed in~\cite{fredricgosfand}. These attacks manipulate the DTM system by subtly altering temperature readings as depicted in Fig.~\ref{fig:threat_model}.

\textbf{Thermal HT Behavior-1} follows a temperature Decrease-Increase pattern that involves the following two manipulation phases:
\begin{enumerate}
    \item \textit{Credit Phase}: The HT purposefully reports lower temperature values to accumulate credits, which are later used in the exploitation phase, as shown in the red area in Fig.~\ref{fig:threat_model}a. The credit phase is not actively running, and it takes time to accumulate enough credit to proceed to the next phase.
    \item \textit{Exploitation Phase}: As depicted in the golden area in Fig.~\ref{fig:threat_model}a, the thermal HT raises the temperature by the same amount, maintaining an apparently normal average.
\end{enumerate}

This attack can result in processor overclocking, delayed cooling activation, and potential thermal damage, ultimately compromising system performance and reliability. 

\textbf{Thermal HT Behavior-2} follows a temperature Increase-Decrease pattern and involves the following two manipulation phases:
\begin{enumerate}
    \item \textit{Credit Phase}: The HT purposefully reports higher temperature values, as shown in the purple area in Fig.~\ref{fig:threat_model}b, to simulate a thermal anomaly, potentially triggering unnecessary system responses. The credit phase is not actively running, and it takes time to accumulate enough credit to proceed to the next phase.
    \item \textit{Exploitation Phase}: The HT reduces the reported temperature by the same amount, as depicted in the red area in Fig.~\ref{fig:threat_model}b, masking its manipulation and maintaining a normal average.
\end{enumerate}
This attack can lead to unnecessary processor throttling, overactivation of the cooling system, and long-term wear on thermal management components, ultimately impacting system efficiency and longevity. 

Both attacks are subtle in nature, exploiting trust in thermal sensor readings, and can result in suboptimal performance, increased power consumption, or hardware damage. As the complexity of multi-core systems increases, these attacks can significantly affect NoC architectures, disrupting inter-tile communication and causing application-level stalling~\cite{elahi2024}. For instance, Tiwari et al.  studied thermal attacks on MPSoCs during multicast operations, where the same information is shared among multiple hardware units through the on-chip network~\cite{Tiwari2019EffectOH}. Detection remains challenging, with recent work focusing on high-sensitivity on-chip temperature sensors\cite{Yan2018ModelingOT}. 
Furthermore, the long-term effects of the aforementioned thermal HTs, such as bias temperature instability (BTI), exacerbate the problem. BTI can alter power density and temperature profiles over time, with studies showing a 5-7\% reduction in maximum power density and a 6-10\% decrease in maximum temperature after 10 years~\cite{sachdeva2023long}. Recent research in FPGA security has also demonstrated the feasibility of remote fault injection attacks that exploit voltage and temperature fluctuations. The \textit{RAM-Jam} attack~\cite{RAMJAM2024} leverages memory collisions in dual-port RAMs to induce transient short circuits, resulting in significant voltage drops and heating. These effects lead to timing violations, bit-flips in configuration memory, and security breaches in FPGA-based systems. Such findings further emphasize the risks posed by temperature-based hardware Trojans, as they can be exploited to accelerate degradation and compromise system integrity over time.

%% file: 3-Methodology.tex
% \vspace{-2mm}
\section{Data Collection Methodology}
\label{sec:method}

Designing an efficient anomaly detection module requires a thorough understanding of the available signals that are triggered when a thermal HT is activated. To this end, we have collected a dataset of MPSoC behavior encompassing the system's performance, thermal, and network parameters. Our novel data collection process integrates widely used simulators: 1) CoMeT~\cite{comet}, an MPSoC performance simulator, to track system's performance at scale; 2) AccessNoxim~\cite{7330124}, a NoC simulator providing cycle-accurate network-on-chip architectural simulations; and 3) HotSpot~\cite{hotspot}, a thermal model that enables precise temperature estimations for various MPSoC layouts. This integrated approach allows comprehensive analysis of performance, power consumption, and thermal behavior in complex MPSoC designs.

The aggregated simulation platform allows recording a comprehensive set of features relevant to the chip's performance as well as NoC performance and security. The proposed simulation is configured to capture data at regular intervals, providing a detailed view of network behavior over time. As listed in Table~\ref{tab:feature_set}, the key parameters collected from the aggregated simulation platform include:

\begin{itemize}
\item Temporal data: Event Cycle, representing the sequence of steps in packet processing and routing.
\item Core metrics: Core Power, Core Temperature, Core Utilization, and Core Frequency, offering insights into processor performance and thermal characteristics.
\item Network traffic data: Packet Source, Packet Destination, Flit Type, Flit Hop Count, Flit Sequence No, and Packet Sequence No, providing detailed information about packet movement through the network.
\item Router-specific data: Current Router, Receiving Port, Departing Port, Router's Congestion, Router's Temperature, Router's Temperature 2-Cycle Average, and Router's Temperature Running Average, capturing the state and performance of individual routers.
\item Security-related data: Attack Label, identifying potential security threats or attacks within the network.
\end{itemize}

The collected data using the aggregated simulation platform has gone through multiple pre-processing steps to ensure data quality and to prepare it for analysis. Step-1) \textit{Data cleaning} step at which we remove any incomplete or erroneous entries from the collected dataset. Step-2) \textit{Feature scaling} step normalizes numerical features to ensure consistent scale across different metrics. Step-3) \textit{Encoding categorical variables} converts categorical features like Flit Type into numerical representations suitable for machine learning algorithms. Step-4) \textit{Temporal alignment} ensures all collected data points are properly synchronized with respect to the Event Cycle. Step-5) \textit{Feature engineering} to create derived features such as the Router's Temperature 2-Cycle Average and Temperature Running Average to capture temporal trends in thermal behavior.

% Redundant This comprehensive data collection and preprocessing approach provides a rich dataset that captures the complex interactions within the NoC, enabling in-depth analysis of network performance, thermal characteristics, and potential security vulnerabilities.

\subsection{Feature Analysis and Selection}

To assess the contribution of each feature to our module's protective performance, we employed a random forest classifier, leveraging its built-in feature importance mechanism. This method evaluates the importance of each feature by measuring the mean decrease in impurity (Gini importance) across all decision trees in the ensemble. Specifically, features contributing more to reducing classification uncertainty receive higher importance scores.

\begin{figure}[t]
  \centering
  %\vspace{-10}
  \includegraphics[width=\columnwidth,height=0.8\textheight,keepaspectratio]{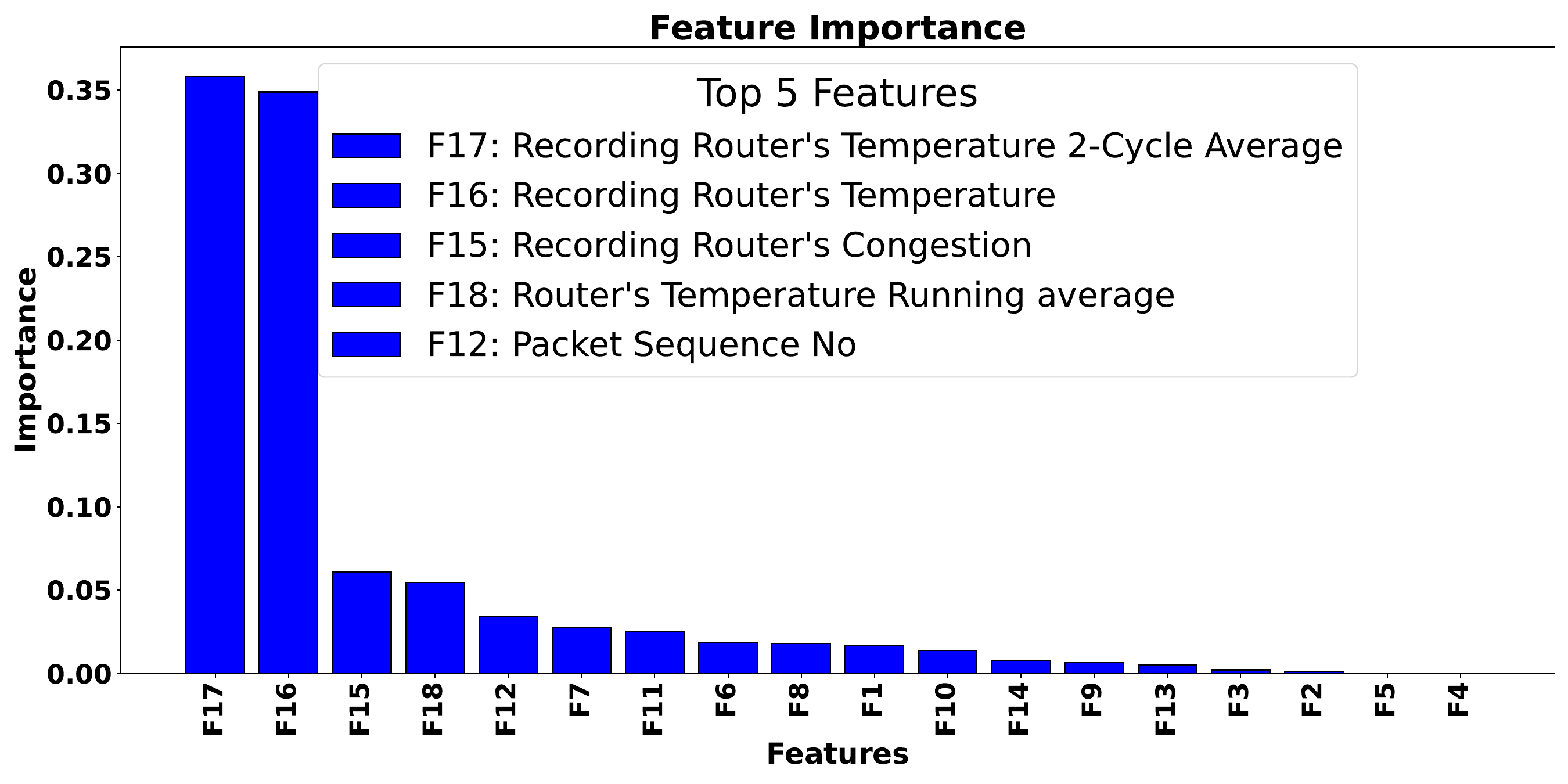}
  \caption{Feature Importance in Network Model: Bar plot showing the importance of features, labeled by their corresponding feature numbers (F1, F2, ..., F19). Higher bars indicate a greater influence on the system's performance.}
  \label{fig:featureimp}
\end{figure}

Based on our feature importance analysis of the recorded parameters (shown in Fig.~\ref{fig:featureimp}), we have identified five distinct combinations of features, each offering unique insights into system performance and thermal characteristics. These feature sets were selected through a combination of empirical analysis and domain-specific knowledge. We first ranked features based on their importance scores obtained from the Random Forest model and then grouped them based on their relevance to key aspects of NoC performance, such as thermal behavior, congestion, and traffic dynamics. The proposed feature sets are defined below.

\begin{enumerate}
    \item \textbf{Thermal-Congestion Set} combines ``Router's Congestion'' (1.69\% importance), ``Router's Temperature 2-Cycle Average'' (35.80\%), and ``Router's Temperature'' (34.88\%). With a combined importance of 72.37\%, this set focuses on the interplay between thermal dynamics and network congestion.

    \item \textbf{Multi-scale Thermal Analysis Set} includes ``Router's Temperature 2-Cycle Average'' (35.80\%), ``Router's Temperature'' (34.88\%), and ``Router's Temperature Running Average'' (5.47\%). This set accounts for 76.15\% of total feature importance. It facilitates a comprehensive analysis of thermal behavior across different time scales.% redundant, which is crucial for understanding the thermal profile of the NoC.

    \item \textbf{Temporal-Thermal Correlation Set} integrates ``Router's Temperature 2-Cycle Average'' (35.80\%), ``Temperature Running Average'' (5.47\%), and ``Event Cycle'' (6.09\%). With a combined importance of 47.36\%, it enables the examination of thermal dynamics in relation to network events and temporal patterns.

    \item \textbf{Network Traffic Characterization Set} consists of ``Temperature Running Average'' (importance not explicitly provided), ``Packet Destination'' (2.79\%), and ``Flit Hop Count'' (1.39\%). This set focuses on network traffic patterns and routing efficiency. While its combined importance is lower, it provides valuable insights into traffic behavior.

    \item \textbf{Thermal-Routing Correlation Set} integrates ``Router's Temperature'' (34.88\%), ``Event Recording Router'' (1.82\%), and ``Packet Source'' (1.85\%). With a combined importance of 38.55\%, it enables the exploration of potential relationships between thermal behavior and packet routing origins.
\end{enumerate}
%\textcolor{red}{Is there any fundamental detail for each set to be explained?} Answer: In previose section we did mention all details about each set. 

The curated feature sets are designed to provide multifaceted perspectives on NoC behavior, with a strong emphasis on thermal characteristics, which our analysis shows to be the most critical factors (\ref{sec:method}). The first three sets, which heavily incorporate thermal features, account for the highest combined importance, reflecting the dominance of thermal dynamics in NoC behavior.

By analyzing these diverse combinations, we aim to uncover complex relationships and patterns within the NoC that may not be apparent when considering features in isolation. This approach facilitates a more nuanced understanding of the interplay between thermal dynamics, network performance, and potential security implications in NoC systems. This refined feature selection, guided by our importance analysis, provides a robust framework for investigating NoC characteristics, with particular emphasis on thermal-related security vulnerabilities and performance implications.

%% file: 4-Proposed.tex
\section{Anomaly Detection Method}
\label{sec:Proposed}

The proposed anomaly detection mechanism in NoC systems employs a statistically-driven approach for monitoring selected features. It defines high and low thresholds for the selected features and routinely checks the number of features out of their ranges to decide on whether or not an abnormal condition. This simple yet efficient method strikes a balance between accuracy and computational efficiency, which is crucial for the resource-constrained NoC environment. Our threshold definitions are based on the following formulas:
\begin{equation}
\begin{cases}
    Threshold_{\text{lower}} = \text{Mean} - K \times \text{std} \\
    Threshold_{\text{upper}} = \text{Mean} + K \times \text{std}
\end{cases}
\label{eq:thresholds}
\end{equation}

where \textit{Mean} represents the average value of a feature, \textit{std} is its standard deviation, and $K$ is an adjustable sensitivity factor. The effectiveness of the proposed formulation and method stems from several key factors:

\begin{itemize}
    \item \textbf{Incorporation of central tendency:} The mean anchors the threshold to typical system behavior, providing a reliable baseline for anomaly detection.
    \item \textbf{Accounting for data spread:} By including the standard deviation (std), the formula adapts to the inherent variability of features, crucial for handling diverse datasets and conditions in NoC environments \cite{zang2018markov,jin2024survey}.
    \item \textbf{Adjustable sensitivity:} The factor $K$ allows fine-tuning of threshold sensitivity, balancing between false positives and missed anomalies. This flexibility is essential in NoC systems where the cost of false alarms must be weighed against the risk of undetected issues~\cite{tang2022rethinking}.

    \item \textbf{Normal distribution assumption:} For normally distributed data, this formula captures specific data percentages within thresholds (e.g., $K=1$ for 68\%, $K=2$ for 95\%, $K=3$ for 99.7\%), providing a statistical foundation for anomaly classification~\cite{5771200}.

    \item \textbf{Contextualizing Threshold Sensitivity:} The confidence levels associated with each threshold ($K=1$ for 68\%, $K=2$ for 95\%, etc.) are based on the assumption of normality in the sampled data. However, although the temperature behavior of the NoC is following the normal distribution but other features may exhibit varying distributions~\cite{mhatre2014temperature}, which could affect the effectiveness of these thresholds. If the distribution deviates significantly from normality (e.g., skewed or multimodal), these fixed confidence intervals may not hold~\cite{pek2017confidence}. To account for this, we assess the empirical distribution of features and adjust $K$ accordingly, ensuring that the thresholds remain representative of actual system behavior~\cite{kelley2005effects}. A detailed discussion of this adaptation is provided in Section~\ref{subsec:stdappr} and~\ref{subsec:AverageApp}.
    
    \item \textbf{Robustness to outliers:} The use of absolute values creates symmetric thresholds, reducing the impact of extreme outliers\textemdash a common challenge in NoC data streams \cite{d2021malicious,esmaeili2023anomaly}.

    \item \textbf{Computational efficiency:} Utilizing simple statistical measures allows for real-time updates, crucial for online anomaly detection in NoCs \cite{putrada2023predictive}.
    \item \textbf{Adaptivity to changing conditions:} The ability to update mean and standard deviation over time ensures the method remains effective as system behavior evolves \cite{wang2021dl,ma2021comprehensive}.
\end{itemize}

Due to the high computational complexity and hardware demands of standard deviation calculations, we developed a series of hardware-efficient approximation techniques for anomaly detection thresholds. These methods balance statistical accuracy with the need for real-time processing in resource-constrained NoC systems.

\subsubsection{STD approximation}
\label{subsec:stdappr}
As discussed, the proposed method uses a multi-tiered thresholding for anomaly detection. The std approximation uses bit-shift operations to calculate various sigmas which are in turn used to achieve various levels of anomaly detection granularity. In fact, we update the thresholds as shown in Eq. \ref{eq:approximatethresholds} below:
\begin{figure}[t]
  \centering
  %\vspace{-10}
  \includegraphics[width= 0.5\textwidth]{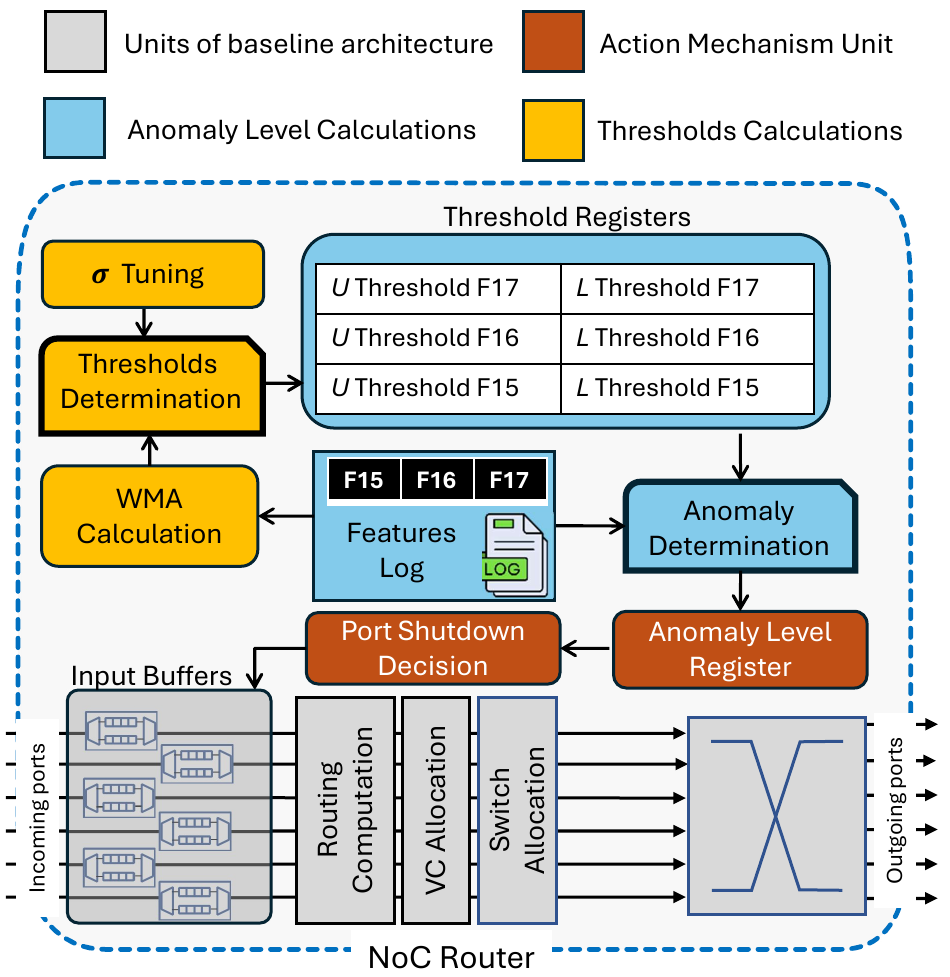}
  \caption{Proposed router architecture based on the anomaly detection and action mechanism modules.}
  \label{fig:routerarch}
\end{figure} 
\begin{equation}
\begin{cases}
    Threshold_{\text{lower}} = \text{Mean} - \sigma_n, \quad 1 \leq n \leq 7, \\
    Threshold_{\text{upper}} = \text{Mean} + \sigma_n, \quad 1 \leq n \leq 7.
\end{cases}
\label{eq:approximatethresholds}
\end{equation}

where $\sigma_n$ is: 
\begin{equation}
\text{$\sigma_n$} = \frac{\text{Mean}}{2^n}, \quad 1 \leq n \leq 7
\label{eq:sigman}
\end{equation}

In this method, $\text{Mean}$ represents the Average of the monitored feature and $\sigma_n$ 
 is a replacement of the standard deviation as an approximation because of the complexity of the implementation of the standard deviation. The use of $2^n$ in the denominator allows for efficient computation using bit-shift operations, which are significantly faster and less resource-intensive than division operations in hardware implementations. This approach aligns with recent trends in energy-efficient edge computing solutions for anomaly detection, such as the TinySNN framework, which emphasizes hardware efficiency in real-time anomaly detection systems~\cite{5771200}. Similarly, the proposed architecture in~\cite{zhang2024mixpequantizationhardwarecodesign} demonstrates the importance of low size, weight, and power (SWaP) machine learning solutions for edge devices, utilizing bit operations to achieve significant power-latency reductions compared to conventional deep networks. Our analysis revealed that the first four thresholds ([$\text{$\sigma_1$}$,\text{$\sigma_4$}]) encompassed 100\% of the data points, rendering them ineffective for anomaly detection. However, the narrower thresholds provided more nuanced results:
\begin{itemize}
    \item $\text{Mean} \pm \text{$\sigma_5$}$: 96.83\% Confidence Interval
    \item $\text{Mean} \pm \text{$\sigma_6$}$: 81.80\% Confidence Interval
    \item $\text{Mean} \pm \text{$\sigma_7$}$: 61.97\% Confidence Interval
\end{itemize}
This tiered approach enables the detection of anomalies across a spectrum of severities, from low to high, offering improved sensitivity compared to broader thresholds.

\subsubsection{Average Approximation}
\label{subsec:AverageApp}
To optimize for the resource constraints of NoC environments, we leveraged bit-shift operations, which are computationally lightweight and efficient in hardware implementations. Furthermore, we introduced a runtime $\text{Mean}$ calculation method using a Weighted Moving Average (WMA) instead of regular $\text{Mean}$:
\begin{equation}
\text{WMA}_1 = \frac{w_1 x_1 + w_0 \text{WMA}_0}{w_1 + w_0}
\label{eq:onlinemean}
\end{equation}
where:
\begin{itemize}
    \item $\text{WMA}_1$ is the new weighted moving average
    \item $x_1$ is the current observation
    \item $\text{WMA}_0$ is the previous weighted moving average
    \item $w_1$ and $w_0$ are weights based on the current temperature $T$. 
\end{itemize}

\begin{figure}[t]
  \centering
  %\vspace{-10}
  \includegraphics[width=0.5\textwidth]{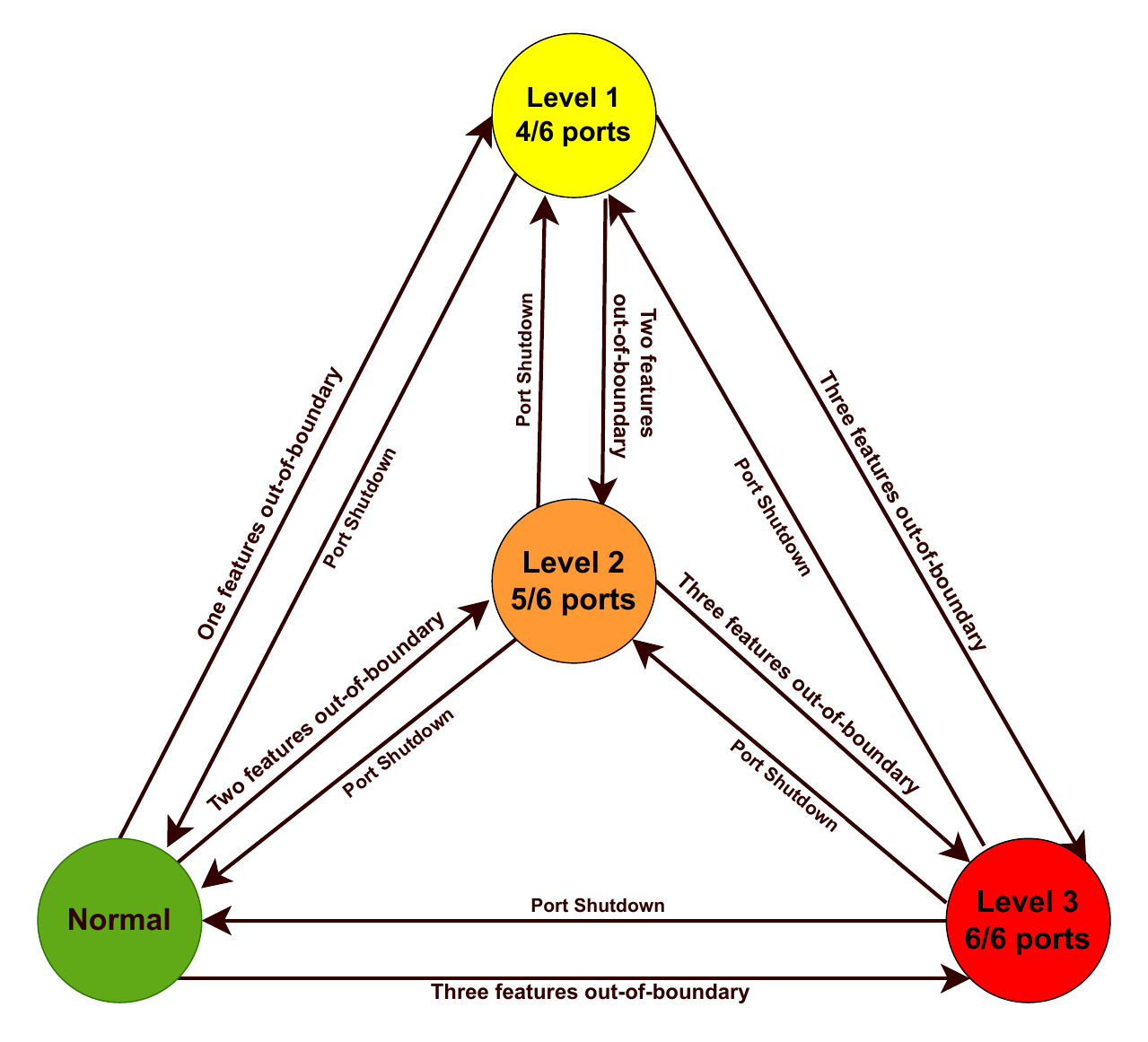}
  \caption{State transition diagram illustrating the Anomaly Detection Module's response to different attack scenarios and corresponding port shutdown states.}
  \label{fig:statetransitiondiagram}
\end{figure}
The weight \( w_i \) is defined as:
\begin{equation}
w_i = 
\begin{cases} 
1, & \text{if } x < T_1 \\ 
3, & \text{if } T_1 < x < T_2 
\end{cases}, \quad i = 0, 1
\label{eq:weightobtain}
\end{equation}

This WMA approach offers several advantages tailored to NoC systems:
\begin{itemize}
    \item Adaptability to recent temperature changes
    \item Memory efficiency, storing only the previous $\text{WMA}$ value
    \item Sensitivity control through adjustment of $T_1$ and $T_2$ thresholds
\end{itemize}

By combining these hardware-level approximation techniques with our multi-tiered thresholding system, we have developed a robust and efficient approach to anomaly detection in NoC environments. This method balances the need for statistical accuracy with the practical constraints of NoC hardware, enabling real-time monitoring and rapid response to potential system anomalies.

By having an approach to measure and find the anomaly we need policies to enhance the security and thermal resilience of NoC systems, we have developed an adaptive real-time monitoring and action mechanism for NoC routers as illustrated in Fig.~\ref{fig:routerarch} which will be discussed in \ref{sec:subaction}.
\subsection{Real-time Monitoring by Anomaly Detection Policies}
\label{sec:subaction}
 This mechanism continuously evaluates three critical features related to temperature and network performance, comparing each against predefined thresholds. The status of each feature is categorized into one of three states: Upper (U), Lower (L), or Normal (N).
 Based on the collective status of these features, we define four distinct operational scenarios: Normal, where all features remain within boundary thresholds, and three levels of anomaly. Anomaly Level 3 occurs when all three features in a chosen set are outside the thresholds boundary. Anomaly Level 2 begins when two of the features in a chosen set are outside the threshold boundary. Anomaly Level 1 happens when one of the features in a chosen set is outside the threshold boundary. Subsequently, the thresholds are calculated and stored in the $\textit{Approximate Thresholds Registers}$. The logged features are then compared with the calculated thresholds, and the results are recorded in the $\textit{Anomaly Level Register}$.
With regard to the captured anomaly level, as shown in Fig.~\ref{fig:routerarch}, each anomaly level triggers a proportional response from the router, carefully balancing network security and thermal management. In our study, we consider port shutdown router scenarios with the following graduated responses. The state transition diagram of the Anomaly Detection Module and its corresponding action mechanisms are shown in Fig.~\ref{fig:statetransitiondiagram}.
\begin{figure*}[t]
  \centering
  %\vspace{-10}
  \includegraphics[width=\textwidth]{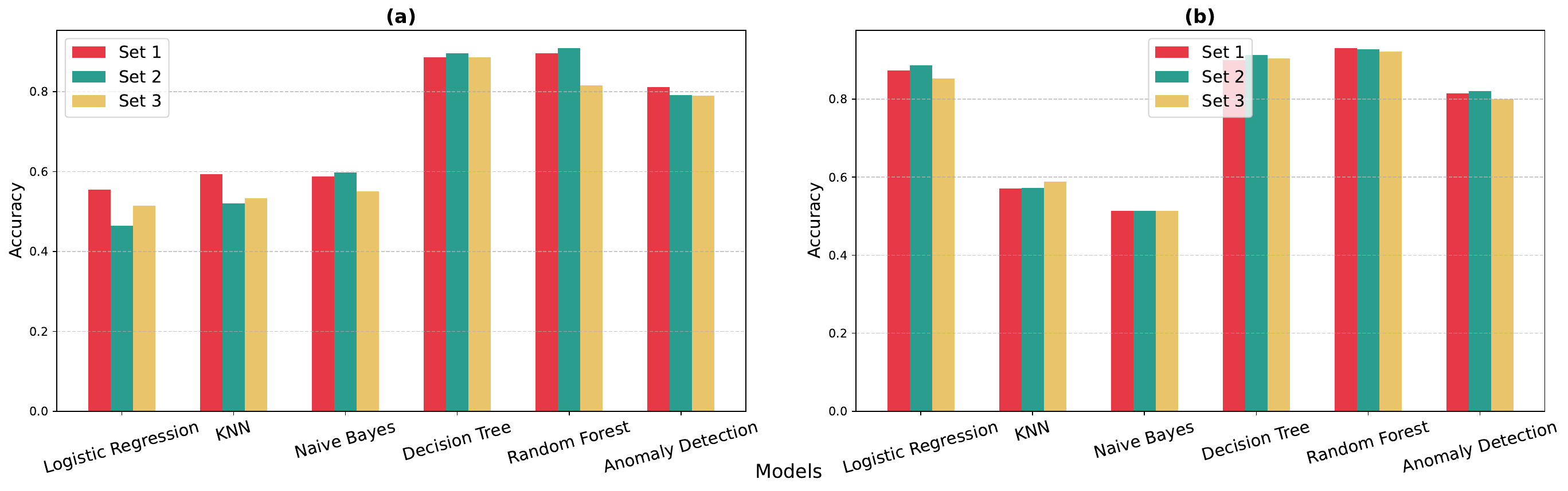}
  \caption{Comparison of classification accuracy between various machine learning models and the proposed anomaly detection module. (a) Binary classification results across multiple sets, including the anomaly detection module's performance. (b) Multi-class classification results across multiple sets, highlighting the anomaly detection module's accuracy alongside other models.}
  \label{fig:ml_comparison}
\end{figure*}
\begin{itemize}
\item \textbf{Anomaly Level 1:}
\begin{itemize}
\item Action: Random shutdown of four ports
\item Impact: Traffic load reduced to 33.33\% of original capacity
\item Outcome: This action provides a significant reduction in thermal load while maintaining minimal network connectivity. The randomness in port selection prevents predictable traffic patterns that attackers might exploit.
\end{itemize}
\item \textbf{Anomaly Level 2:}
\begin{itemize}
\item Action: Random shutdown of five ports
\item Impact: Traffic load reduced to 16.67\% of original capacity
\item Outcome: This more aggressive response is warranted by the higher anomaly level. It drastically reduces thermal load and potential attack surfaces while still allowing critical communications through the remaining port.
\end{itemize}
\item \textbf{Anomaly Level 3:}
\begin{itemize}
\item Action: Complete shutdown of all six ports
\item Impact: Total cessation of traffic through the router
\item Outcome: This extreme measure is justified by the severity of the anomaly. It completely isolates the potentially compromised router, preventing any further damage or attack propagation through the network.
\end{itemize}
\end{itemize}
These action mechanisms are designed to provide a graduated response to detected anomalies, balancing the need for security and thermal management with the maintenance of network functionality. The rationale behind each action is as follows:
\begin{itemize}
\item \textbf{Proportional Response}: By tailoring the action to the severity of the detected anomaly, we avoid overreacting to minor issues while ensuring robust protection against severe threats.
\item \textbf{Thermal Management}: The progressive reduction in active ports directly correlates with a decrease in heat generation, addressing potential thermal attacks or malfunctions effectively.
\item \textbf{Network Resilience}: By maintaining some level of connectivity in lower anomaly levels, we allow the network to potentially self-heal or reroute critical traffic, enhancing overall system resilience.
\item \textbf{Attack Mitigation}: The randomness in port shutdown for Levels 1 and 2 makes it difficult for attackers to predict and exploit traffic patterns, adding an extra layer of security.
\item \textbf{Resource Optimization}: This approach optimizes the use of network resources by only shutting down as many ports as necessary, balancing security needs with performance requirements.
\end{itemize}

This graduated response system provides a nuanced approach to maintaining network security and thermal stability. As the severity of detected anomalies increases, the system implements progressively more aggressive measures to mitigate potential threats and thermal issues. By employing this real-time monitoring and action mechanism, we aim to significantly enhance the robustness and reliability of NoC systems while minimizing unnecessary disruptions to network connectivity.
The proposed method not only tackles urgent security issues but also aids in the ongoing thermal management of the NoC infrastructure, helping to prevent future problems caused by stress, overload, or overheating. By dynamically adjusting the router's operation based on instantaneous data, the system can effectively respond to both sudden security threats and gradual thermal buildup. This adaptive strategy allows for an efficient use of network resources, potentially extending the lifespan of the hardware. Furthermore, this mechanism provides a framework for implementing more sophisticated response strategies in the future. For instance, further enhancing the system's ability to maintain optimal performance under varying conditions. 

\begin{figure*}[t]
  \centering
  %\vspace{-10}
  \includegraphics[width=\textwidth]{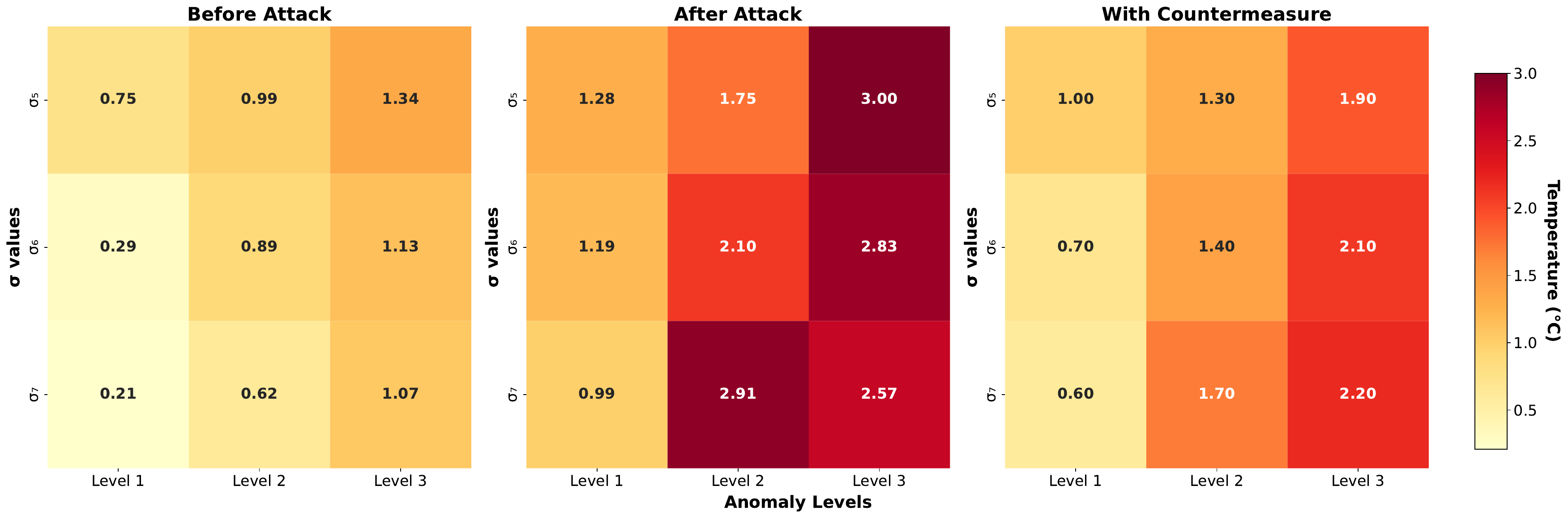}
  \caption{Heatmap visualization of temperature fluctuation averages under different scenarios and anomaly levels. (a) Temperature fluctuations before the attack, showing baseline conditions. (b) Temperature fluctuations after the attack, demonstrating the impact of the anomaly. (c) Temperature fluctuations with countermeasures applied, illustrating the effectiveness of mitigation strategies. Each heatmap displays data for three sigma values ($\sigma_5$, $\sigma_6$, $\sigma_7$) across three anomaly levels.}
  \label{fig:tempfluc}
\end{figure*} 
\subsection{Approximation-Based Anomaly Detection Module Architecture}

The proposed anomaly detection module architecture, as illustrated in Fig.~\ref{fig:routerarch}, employs several key components to efficiently detect and respond to anomalies in NoC systems. At the core of the architecture is the \textit{Features Log}, which stores recent values of monitored features, providing essential input data for the detection process. The \textit{Online Approximate Weighted Moving Average Module} processes this logged data to compute an approximation of the moving average of the features, offering a computationally efficient method for tracking feature trends over time.

The $\sigma$ \textit{Tuning} component plays a crucial role in determining the appropriate $\sigma_n$ value, which is used to adjust the sensitivity of the anomaly detection thresholds. This adaptive approach allows the system to remain responsive to changing network conditions and potential threats. To simplify the implementation of the $\Sigma$ calculation, it is performed using logical right shifts at the hardware level. Calculated threshold values are stored in the \textit{Approximate Threshold Registers}, which are continuously updated based on the outputs from the \textit{Online Approximate Weighted Moving Average} and \textit{$\sigma$ Tuning} components.

The comparison between current feature values from the \textit{Features Log} and the thresholds stored in the \textit{Approximate Threshold Registers} is a critical step in the anomaly detection process. The results of this comparison are captured in the \textit{Anomaly Level Register}, which effectively quantifies the degree of deviation from normal behavior. Finally, the \textit{Port Shutdown Decision Module} interprets the anomaly level recorded in the \textit{Anomaly Level Register} and determines the appropriate action to take, which may include shutting down a number of ports to mitigate potential threats or thermal issues.

The anomaly detection process flows seamlessly through these components, starting with the continuous logging of feature values in the \textit{Features Log}. The \textit{Online Approximate Weighted Moving Average Module} then processes these logged features, while the $\sigma$ \textit{Tuning} component determines the appropriate $\sigma_n$ value. New thresholds are calculated using Eq.~\ref{eq:approximatethresholds}, and the \textit{Approximate Threshold Registers} are updated accordingly. Current feature values are then compared against these thresholds, with the results recorded in the \textit{Anomaly Level Register}. Based on the anomaly level indicated in the \textit{Anomaly Level Register}, the \textit{Port Shutdown Decision Module} initiates the appropriate response.

This architecture enables real-time anomaly detection and response while minimizing computational overhead, making it particularly suitable for implementation in NoC routers. The use of approximation techniques in the WMA calculation and threshold determination allows for efficient hardware implementation without significantly compromising detection accuracy. By integrating these components, the proposed architecture provides a robust and adaptive solution for maintaining the security and performance of NoC systems in the face of potential anomalies and attacks.

%% file: 5-Results.tex
\section{Experimental Results and Analysis}
\label{sec:results}

This study employs a two-stage simulation approach, utilizing a Sniper-based simulator~\cite{sniper}, CoMeT~\cite{comet}, for traffic generation and core-memory feature extraction, along with AccessNoxim for NoC simulation~\cite{7330124}. This setup enables the evaluation of real-world benchmark performance in a NoC environment by integrating both core-memory and network transactions. To the best of our knowledge, this work is the first to provide a comprehensive evaluation platform that combines both CoMeT and AccessNoxim, enabling the simultaneous analysis of core-memory and network transactions. This integration grants access to core-memory temperature and performance metrics for real-world benchmarks while also capturing network performance and temperature characteristics. This dual-perspective approach allows for a more holistic assessment of MPSoC behavior under realistic workloads.  

\begin{figure*}[t]
  \centering
  %\vspace{-10}
  \includegraphics[width=\textwidth]{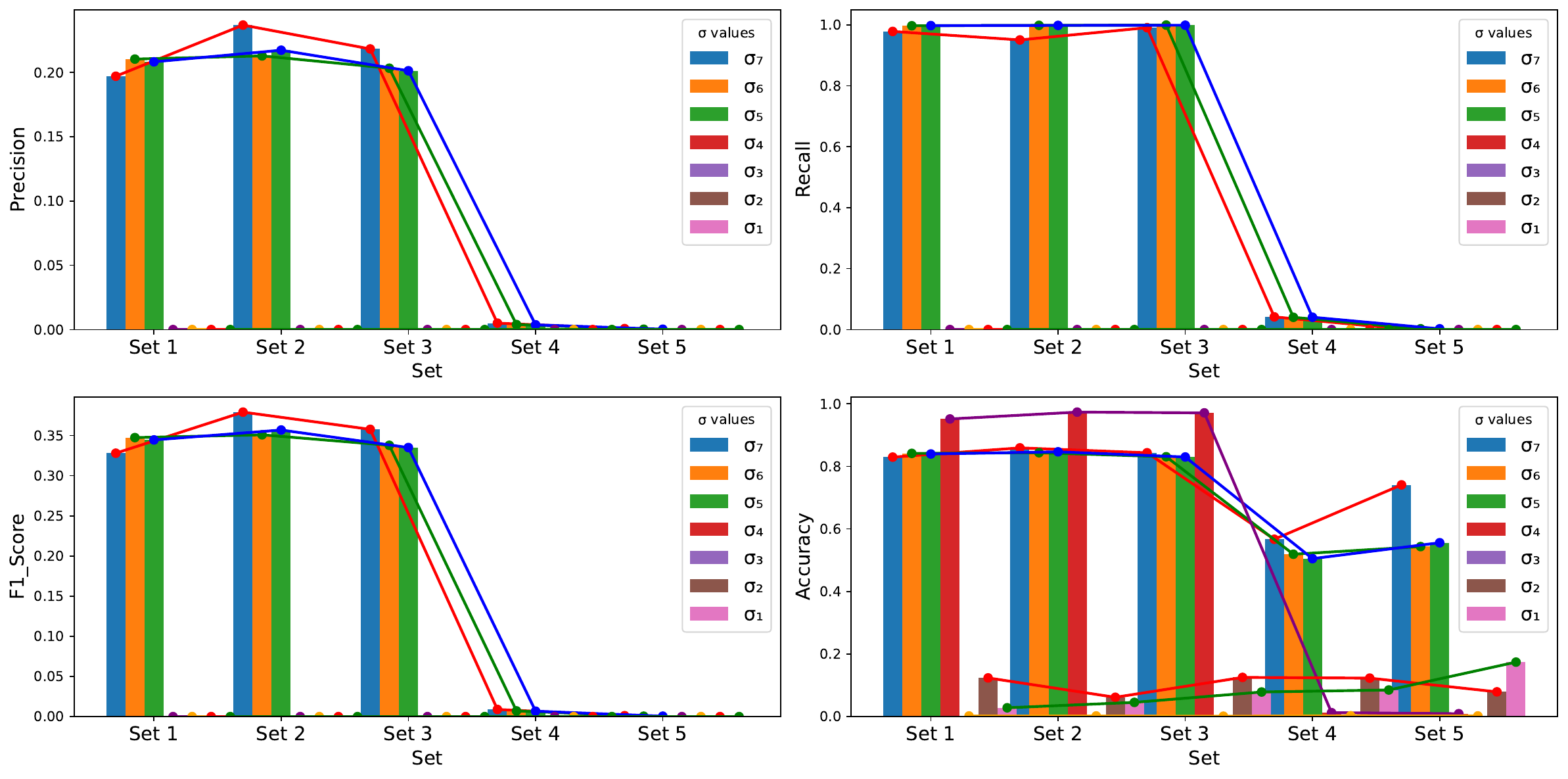}
  \caption{Performance comparison of anomaly detection module metrics (Precision, Recall, F1 Score, and Accuracy) across different feature sets (Set 1 to Set 5) and sigma thresholds ($\sigma_7$ to $\sigma_1$). Panels (a), (b), (c), and (d) present the metrics for each feature set, with each subplot showing the variations of the respective metric (Precision, Recall, F1 Score, and Accuracy) as sigma values decrease.}
  \label{fig:metric_comparison}
\end{figure*} 
CoMeT (Core and Memory Thermal Simulator) is an advanced, integrated interval thermal simulation toolchain designed for various processor-memory systems. In this study, CoMeT is used specifically to export real-world benchmark network traffic. It supports multiple core-memory configurations, including 2D, 2.5D, and 3D architectures. Additionally, it provides detailed performance metrics at user-defined intervals and offers an extensive API for data export and manipulation.  

AccessNoxim, an enhanced version of the Noxim NoC simulator \cite{7330124}, is utilized to simulate the network using the traffic data exported from CoMeT. It supports various network topologies, including both direct and indirect wired networks, and allows for configurable parameters such as network size, buffer size, packet size distribution, and routing algorithms. Additionally, it enables the import of custom traffic patterns through table-based traffic distribution.  

By leveraging this integrated simulation environment, we have collected 19 features related to both the MPSoC and the corresponding NoC, making this the first study of its kind. These features provide a detailed characterization of performance and thermal behavior across different system layers, contributing to a deeper understanding of NoC security and efficiency.  

This simulation environment enables the analysis of real-world benchmark performance in a NoC context, providing insights into network behavior under realistic workloads. The integration of CoMeT-generated traffic into AccessNoxim allows for a more accurate representation of actual system behavior, enhancing the validity of the simulation results. In this simulation setup, the network traffic is gathered using real-world benchmarks from CoMeT and subsequently processed into the AccessNoxim simulator, incorporating uniform background traffic. The results in this section pertain to the PARSEC benchmarks~\cite{parsec}, implemented with 5\% background uniform traffic.
Moreover, HT was integrated into the router to manipulate the temperature sensor as outlined in the attack model in Section~\ref{sec:threat}. Throughout the simulations, attacks are temporally distributed according to a normal distribution and spatially distributed in a uniform manner.

\subsection{Performance Comparison of Feature Sets}

Our analysis of different sigma thresholds across five feature sets revealed significant variations in model performance, emphasizing the importance of careful feature selection and threshold calibration in anomaly detection systems. The performance metrics presented in  Figs.~\ref{fig:metric_comparison}a, b, c, and d were obtained from our implemented anomaly detection module, which operates at the network level. This implementation allows for a more realistic, large-scale evaluation of our approach, where anomalies can occur at various points in a complex network environment. By testing the model on these large-scale scenarios, we can assess the scalability and robustness of the anomaly detection system.

\subsubsection{High-Performance Sets}
Feature Sets 1, 2, and 3 consistently demonstrated strong performance across multiple evaluation metrics, particularly with higher sigma values ($\sigma_7$, $\sigma_6$, $\sigma_5$). These sets were able to effectively identify true anomalies with minimal false positives, as indicated by their high precision (ranging from 24\% to 79\%) and very high recall values (greater than 95\%). High recall values imply that the model was highly sensitive to detecting anomalies, ensuring that most of the true anomalies were captured. The strong F1 scores (ranging from 39\% to 88\%) further emphasize the model's ability to maintain a good balance between precision and recall, which is essential in real-world applications where both false positives and false negatives are costly.

Additionally, accuracy remained consistently high (greater than 86\%) across these feature sets, further supporting the effectiveness of the model in identifying anomalies while maintaining a reasonable level of performance in correctly classifying normal instances. These trends were particularly evident in the precision and recall metrics shown in Figs.~\ref{fig:metric_comparison}a and \ref{fig:metric_comparison}b. The overall performance of Feature Sets 1, 2, and 3 suggests that the selected features in these sets capture relevant and distinguishing patterns for anomaly detection. These patterns allow the model to reliably differentiate between normal and anomalous behavior in a network environment, making them ideal candidates for accurate anomaly detection in a large-scale setting.

\subsubsection{Challenging Sets}

In contrast, Feature Sets 4 and 5 posed significant challenges for the anomaly detection system, regardless of the sigma thresholds applied. For these sets, the precision and recall metrics were extremely low, with values falling below 0.09\% and 0.05\%, respectively, even under the best-performing sigma values. These low values indicate that the model struggled to identify true anomalies and produced a large number of false negatives. The poor F1 scores (less than 6\%) further reflected this issue, suggesting that the model was unable to find an acceptable balance between true positives and false positives. These results also indicate that the features in Sets 4 and 5 may not be sufficiently informative or discriminative for effective anomaly detection, making it difficult for the model to reliably classify anomalies.

Although accuracy remained relatively high ($>$86\%) across these sets, this metric proved misleading due to its susceptibility to class imbalance. In cases of imbalanced data, accuracy alone cannot reliably indicate model performance, as the model could achieve high accuracy by simply classifying the majority class correctly, while neglecting the detection of the minority class (in this case, anomalies). The stark performance disparity between Feature Sets 1-3 and Feature Sets 4-5 suggests that the latter may contain fundamentally different data characteristics or noise patterns, which the current model is not equipped to handle effectively. These challenges are clearly evident in the performance metrics displayed in Figs.~\ref{fig:metric_comparison}c and \ref{fig:metric_comparison}d, where the low precision and recall values for Sets 4 and 5 contrast sharply with the performance of the other feature sets.

\subsubsection{Threshold Sensitivity}
\begin{figure*}[t]
  \centering
  %\vspace{-10}
  \includegraphics[width=\textwidth]{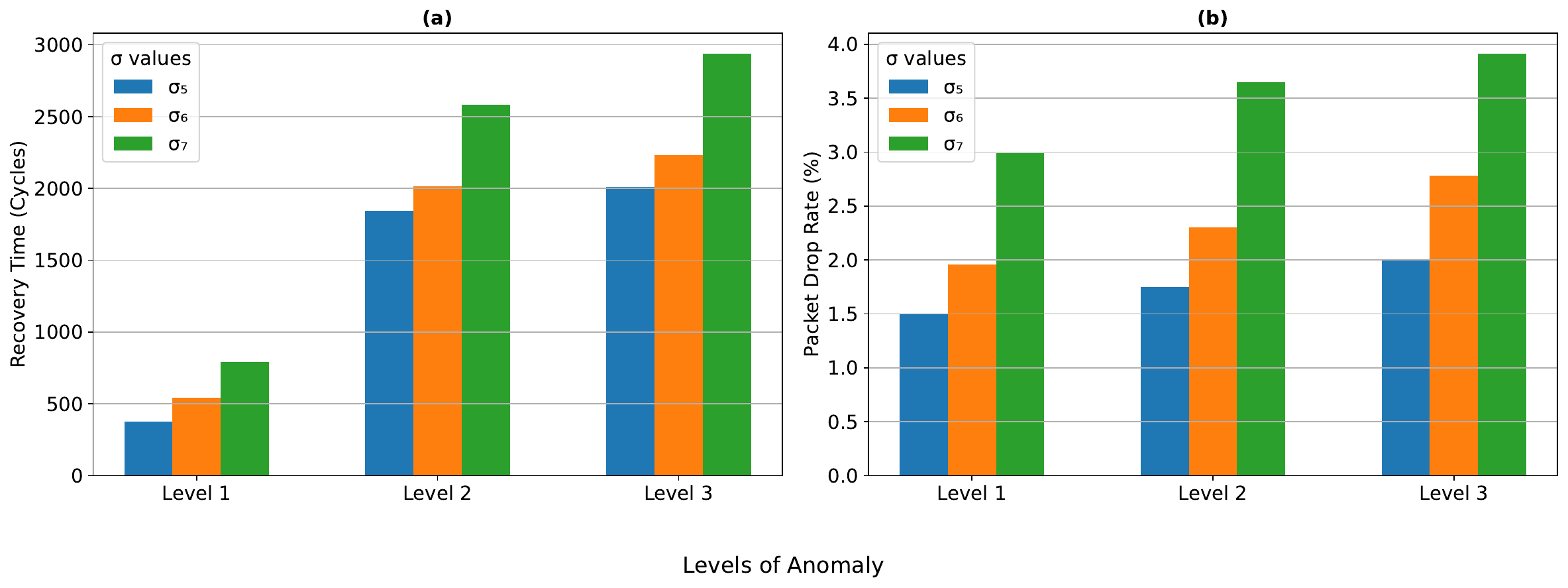}
  \caption{System recovery time and packet drop rate under different anomaly levels and sigma values. (a) Recovery time in cycles for various sigma values ($\sigma_5$, $\sigma_6$, $\sigma_7$) across three anomaly levels. (b) Packet drop rate percentage for the same sigma values and anomaly levels, illustrating the impact on network performance.}
  \label{fig:recoverydroprate}
\end{figure*} 
Our analysis also highlighted the significant sensitivity of the anomaly detection system to the sigma threshold values. As sigma values decreased from $\sigma_7$ to $\sigma_1$, we observed a clear trend of declining performance across all feature sets. Higher sigma values ($\sigma_7$, $\sigma_6$, $\sigma_5$) consistently outperformed the lower sigma values in all evaluation metrics, demonstrating that a more lenient threshold allows the model to capture more anomalies without overfitting to noise. This behavior was particularly important for Feature Sets 1-3, where the higher sigma values enabled the model to achieve strong precision, recall, and F1 scores.

In contrast, the use of lower sigma values, particularly $\sigma_4$, led to a dramatic drop in performance. For Feature Sets 4 and 5, this drop was particularly severe, with zero true positives identified at $\sigma_4$. This suggests that the threshold was too restrictive for these feature sets, failing to detect any anomalies at all. Further decreases in the sigma values (to $\sigma_3$, $\sigma_2$, and $\sigma_1$) resulted in extremely poor performance across all sets, with precision, recall, and F1 scores all nearing zero. These results suggest that overly restrictive thresholds are unsuitable for anomaly detection and may lead to an excessive number of false negatives.

The sensitivity of the model to the choice of sigma values underscores the critical importance of threshold selection in anomaly detection systems. These findings suggest that adaptive thresholding techniques could be highly beneficial for handling diverse feature sets. By dynamically adjusting the threshold based on the characteristics of the data, such techniques could improve the model’s robustness and ability to effectively detect anomalies across a variety of feature sets, regardless of their inherent complexity or noise.

\subsection{Comparison with Machine Learning Models}

Our analysis of various classification models for anomaly detection in NoC systems reveals interesting trade-offs between accuracy and implementation feasibility. As shown in Figs. \ref{fig:ml_comparison}a and b, we compared the performance of several machine learning models against our proposed Anomaly Detection Module across three different datasets for both binary and multi-class classification tasks. 

For our comparison, we evaluated five machine learning models for anomaly detection in NoC systems: Logistic Regression (LR), K-Nearest Neighbors (KNN), Naïve Bayes (NB), Decision Tree (DT), and Random Forest (RF). Additionally, we assessed our Anomaly Detection Module, which is designed for efficient real-time deployment in NoC environments. 

Logistic Regression (LR) is a statistical model that applies a sigmoid activation function to estimate the probability of a given input belonging to a particular class. We used L2 regularization (Ridge) with a regularization strength of \( C = 1.0 \) to prevent overfitting. K-Nearest Neighbors (KNN) is a distance-based algorithm that classifies data points based on the majority label of their \( k \) nearest neighbors. We set \( k = 5 \) and used the Euclidean distance metric for classification. 

Naïve Bayes (NB) is a probabilistic model based on Bayes' theorem, assuming conditional independence between features. We used the Gaussian Naïve Bayes variant, which is suitable for continuous data distributions. Decision Tree (DT) is a tree-based model that recursively splits the dataset based on feature thresholds to minimize entropy, using the Gini impurity criterion. We set the maximum tree depth to 20 to balance accuracy and computational efficiency. 

Random Forest (RF) is an ensemble model consisting of 100 decision trees, trained using bootstrap sampling. The number of features considered for splitting at each node was set to the square root of the total number of features. 

In the binary classification scenario (Fig. \ref{fig:ml_comparison}a), the Random Forest model achieved the highest accuracy, ranging from 81.60\% to 90.81\% across the three datasets. Our Anomaly Detection Module demonstrated competitive performance, with accuracies between 79.04\% and 81.13\%. Similarly, for multi-class classification (Fig. \ref{fig:ml_comparison}b), the Random Forest model again outperformed others with accuracies between 92.17\% and 93.00\%, while our Anomaly Detection Module maintained solid performance with accuracies ranging from 80.04\% to 82.12\%.

While machine learning models, particularly Random Forest, show superior accuracy, it's crucial to consider the practical constraints of NoC environments. Implementing complex ML models in every router would incur significant overhead in terms of computational resources and power consumption. In contrast, our Anomaly Detection Module offers a more lightweight solution that can be feasibly implemented in individual routers.

The Anomaly Detection Module's performance, though slightly lower than the best ML models, remains robust and consistent across different datasets and classification tasks. It achieves accuracies within 10-15\% of the top-performing Random Forest model while offering a much more practical implementation for resource-constrained NoC routers.

This comparison demonstrates that while ML models can provide high accuracy in anomaly detection, our proposed Anomaly Detection Module offers a compelling balance between performance and implementability. It provides a viable solution for real-time anomaly detection in NoC systems, where resource efficiency is as critical as detection accuracy.

\subsection{Anomaly Detection Module Evaluation}

Our anomaly detection system's performance was evaluated across three key metrics: temperature fluctuations, recovery time, and packet drop rate. These metrics were analyzed for different threshold values ($\sigma_5$, $\sigma_6$, $\sigma_7$) and anomaly levels (Level 1, 2, 3).

Fig. \ref{fig:tempfluc} illustrates the temperature fluctuations before and after the attack, as well as with our countermeasure in place. Before the attack, temperature variations were minimal across all threshold values and anomaly levels, ranging from 0.21°C to 1.34°C (Fig. \ref{fig:tempfluc}a). The attack significantly increased these fluctuations, with the highest recorded at 3.00°C for $\sigma_5$ at Level 3 (Fig. \ref{fig:tempfluc}b). Our countermeasure effectively mitigated these fluctuations, reducing them to a range of 0.6°C to 2.2°C (Fig. \ref{fig:tempfluc}c).

The effectiveness of our countermeasure is particularly evident for higher anomaly levels. For instance, at Level 3, the countermeasure reduced the temperature fluctuation from 3.00°C to 1.9°C for $\sigma_5$, from 2.83°C to 2.1°C for $\sigma_6$, and from 2.57°C to 2.2°C for $\sigma_7$. This demonstrates the robustness of our approach across different threshold values.

Fig. \ref{fig:recoverydroprate} presents the recovery time and packet drop rate for each anomaly level and threshold value. The \textit{recovery time} refers to the duration required for the router to return to a stable or nominal state after experiencing an anomaly. In the context of NoC security and thermal resilience, recovery time represents how quickly the components can mitigate the effects of a thermal attack. As shown in Fig. \ref{fig:recoverydroprate}a recovery time increases with both anomaly level and threshold value. For $\sigma_5$, recovery time ranges from 378 to 2009 cycles, while for $\sigma_7$, it extends from 790 to 2936 cycles. This trend indicates that higher threshold values, while potentially reducing false positives, may lead to longer system recovery times when anomalies are detected.
The packet drop rate, illustrated in Fig. \ref{fig:recoverydroprate}b, shows a similar pattern. It increases with both anomaly level and threshold value, ranging from 1.50\% to 2.00\% for $\sigma_5$ and from 2.99\% to 3.91\% for $\sigma_7$. This suggests that while higher thresholds may offer more conservative anomaly detection, they also risk higher packet loss when anomalies occur.

These results highlight the trade-off between sensitivity and system performance in our anomaly detection approach. Lower threshold values ($\sigma_5$) offer quicker recovery and lower packet drop rates but may be more prone to false positives. Higher thresholds ($\sigma_7$) potentially reduce false alarms but at the cost of longer recovery times and higher packet loss during anomalies. Our anomaly detection system demonstrates effective mitigation of temperature fluctuations across all tested scenarios. The choice of threshold value should be carefully considered based on the specific requirements of the NoC system, balancing between rapid response and minimizing false positives.

\begin{table}[ht]
\centering
\caption{Resource Usage for Exact and Approximation Thresholding on Spartan-3 FPGA device.}
\label{tab:resource_usage}
\renewcommand{\arraystretch}{1.3} % Adjust row height
\setlength{\tabcolsep}{6pt} % Adjust column spacing
\begin{tabular}{|l|c|c|c|c|}
\hline
\rowcolor[HTML]{EAF2F8} 
\textbf{Resources} & \textbf{Exact Thresholding} & \textbf{$\boldsymbol{\sigma_7}$} & \textbf{$\boldsymbol{\sigma_6}$} & \textbf{$\boldsymbol{\sigma_5}$} \\ \hline
\textbf{Slice Flip Flops} & 1\% & 0\% & 0\% & 0\% \\ \hline
\textbf{4-input LUTs} & 3\% & 1\% & 1\% & 1\% \\ \hline
\textbf{Occupied Slices} & 4\% & 1\% & 1\% & 1\% \\ \hline
\textbf{Total LUTs} & 4\% & 1\% & 1\% & 1\% \\ \hline
\textbf{Bonded IOBs} & 39\% & 9\% & 9\% & 9\% \\ \hline
\textbf{BUFGMUXs} & 4\% & 0\% & 0\% & 0\% \\ \hline
\textbf{DSP48As} & 20\% & 0\% & 0\% & 0\% \\ \hline
\textbf{Avg. Fanout} & 2.73 & 2.30 & 2.28 & 2.25 \\ \hline
\end{tabular}
\end{table}

\subsection{Hardware Overhead Analysis}

For hardware evaluation, we implemented our design on a Xilinx Spartan-3 FPGA device. We compare the resource utilization of the exact thresholding method against three approximation methods ($\sigma_7$, $\sigma_6$, and $\sigma_5$), as summarized in Table \ref{tab:resource_usage}. The results reveal substantial resource savings achieved through approximation, which has significant implications for resource-constrained applications, particularly in NoC router designs.

The approximation methods demonstrate remarkable efficiency in resource utilization compared to the exact thresholding method. For logic resources, the approximations reduce usage of 4 input LUTs and Occupied Slices from 3-4\% to just 1\%, representing a 66-75\% reduction. This substantial saving in logic resources can lead to more compact NoC router designs, potentially enabling higher router densities or freeing up chip area for other components.

The savings in specialized resources are even more pronounced. The approximation methods completely eliminate the need for BUFGMUXs and DSP48As, which the exact method utilizes at 4\% and 20\% respectively. This 100\% reduction in specialized resource usage is particularly significant for NoC routers, where DSP resources are often limited and valuable. The elimination of DSP usage not only conserves these critical resources but also potentially leads to substantial power savings, a crucial factor in NoC design.

In terms of I/O resources, the approximation methods achieve a dramatic reduction in Bonded IOB usage, from 39\% for the exact method to just 9\% for all approximation variants. This 76.9\% decrease in I/O utilization is crucial for NoC router designs, potentially enabling more efficient communication interfaces within the network.

Even in performance indicators, the approximation methods show improvements. The Average Fanout of Non-Clock Nets decreases from 2.73 for exact thresholding to as low as 2.25 for the $\sigma_5$ approximation. In NoC routers, this reduction in fanout can contribute to better timing performance and further reductions in power consumption, which is critical for overall system efficiency.

These significant reductions make a compelling case for adopting approximation methods in NoC router designs. The massive resource savings can enable more complex router architectures, reduce power consumption, and potentially allow for more efficient NoC implementations. In the context of NoCs, these savings could translate to improved congestion detection, more efficient adaptive routing decisions, or enhanced power management within routers.

While the exact thresholding method may offer higher precision, the considerable resource overhead it incurs may be prohibitive in many NoC applications, where router efficiency directly impacts overall system performance. The approximation methods, with their dramatic resource savings and only slight compromise in precision, present an attractive alternative for resource-constrained NoC designs or those prioritizing power efficiency. This analysis clearly demonstrates that approximation techniques can significantly optimize hardware utilization in threshold calculations. When applied to NoC routers, this approach aligns well with the ongoing research in NoC design. 

%% file: 6-Conclusions.tex
%\vspace{-2mm}
\section{Conclusions}
\label{sec:Conclusions}
%\textcolor{red}{needs to be enriched}
This study introduces a two-stage simulation environment utilizing the Sniper and AccessNoxim simulators to assess NoC performance using real-world benchmarks. By combining CoMeT for core-memory feature extraction and AccessNoxim for network simulation, the study provides insights into both core-memory and network transaction performance, particularly focusing on temperature metrics and performance under realistic workloads. The research highlights the effectiveness of different feature sets in anomaly detection systems, showing that feature sets 1-3 significantly outperform sets 4-5 in terms of precision, recall, and F1 scores. It emphasizes the importance of selecting appropriate sigma thresholds, as higher thresholds generally deliver better anomaly detection performance by accommodating noise without overfitting. The study also compares various machine learning models, like Random Forest, with a tailored Anomaly Detection Module for implementation in NoC systems. While Random Forest achieves higher accuracy, the proposed module offers a more feasible solution due to its lightweight nature suitable for deployment in resource-constrained environments. Temperature fluctuations, recovery times, and packet drop rates are used to evaluate the anomaly detection system's effectiveness. The trade-off between sensitivity and performance is addressed, with higher thresholds leading to longer recovery times and higher packet loss but reducing false positives. In terms of hardware implementation, the study demonstrates substantial resource savings using approximation methods over exact thresholding in an FPGA setting, especially concerning DSP and I/O resources. This approach not only conserves resources but also enhances power efficiency and performance, making it advantageous for NoC router designs.

\section{Acknowledgment}
This work was supported in part by the National Science Foundation under grants Nos. 2302537 and 2219679.

%% file: main.bbl
\begin{thebibliography}{10}
\providecommand{\url}[1]{#1}
\csname url@samestyle\endcsname
\providecommand{\newblock}{\relax}
\providecommand{\bibinfo}[2]{#2}
\providecommand{\BIBentrySTDinterwordspacing}{\spaceskip=0pt\relax}
\providecommand{\BIBentryALTinterwordstretchfactor}{4}
\providecommand{\BIBentryALTinterwordspacing}{\spaceskip=\fontdimen2\font plus
\BIBentryALTinterwordstretchfactor\fontdimen3\font minus \fontdimen4\font\relax}
\providecommand{\BIBforeignlanguage}[2]{{%
\expandafter\ifx\csname l@#1\endcsname\relax
\typeout{** WARNING: IEEEtran.bst: No hyphenation pattern has been}%
\typeout{** loaded for the language `#1'. Using the pattern for}%
\typeout{** the default language instead.}%
\else
\language=\csname l@#1\endcsname
\fi
#2}}
\providecommand{\BIBdecl}{\relax}
\BIBdecl

\bibitem{Yan2018ModelingOT}
\BIBentryALTinterwordspacing
M.~Yan, H.~Wei, and M.~Onabajo, ``Modeling of thermal coupling and temperature sensor circuit design considerations for hardware trojan detection,'' \emph{2018 IEEE 61st International Midwest Symposium on Circuits and Systems (MWSCAS)}, pp. 857--860, 2018. [Online]. Available: \url{https://api.semanticscholar.org/CorpusID:59234712}
\BIBentrySTDinterwordspacing

\bibitem{Cheng2005HeterogeneousMS}
\BIBentryALTinterwordspacing
X.~Cheng, ``Heterogeneous multi-processor soc: An emerging paradigm of embedded system design and its challenges,'' in \emph{International Conference on Embedded Software and Systems}, 2005. [Online]. Available: \url{https://api.semanticscholar.org/CorpusID:30594962}
\BIBentrySTDinterwordspacing

\bibitem{RAMJAM2024}
M.~M. Alam, S.~Tajik, F.~Ganji, M.~Tehranipoor, and D.~Forte, ``Ram-jam: Remote temperature and voltage fault attack on fpgas using memory collisions,'' in \emph{2019 Workshop on Fault Diagnosis and Tolerance in Cryptography (FDTC)}, 2019, pp. 48--55.

\bibitem{Trop2024}
\BIBentryALTinterwordspacing
S.~Sankar, R.~Gupta, J.~Jose, and S.~Nandi, ``Trop: Trust-aware opportunistic routing in noc with hardware trojans,'' \emph{ACM Trans. Des. Autom. Electron. Syst.}, vol.~29, no.~2, Feb. 2024. [Online]. Available: \url{https://doi.org/10.1145/3639821}
\BIBentrySTDinterwordspacing

\bibitem{HardwareTrojanDetection2023}
\BIBentryALTinterwordspacing
H.~Wang and B.~Halak, ``Hardware trojan detection and high-precision localization in noc-based mpsoc using machine learning,'' in \emph{Proceedings of the 28th Asia and South Pacific Design Automation Conference}, ser. ASPDAC '23.\hskip 1em plus 0.5em minus 0.4em\relax New York, NY, USA: Association for Computing Machinery, 2023, p. 516–521. [Online]. Available: \url{https://doi.org/10.1145/3566097.3567922}
\BIBentrySTDinterwordspacing

\bibitem{huang2024}
A.~Z. Benelhaouare, I.~Mellal, M.~Oumlaz, and A.~Lakhssassi, ``Mitigating thermal side-channel vulnerabilities in fpga-based sip systems through advanced thermal management and security integration using thermal digital twin (tdt) technology,'' \emph{Electronics}, vol.~13, no.~21, 2024.

\bibitem{MitigationHardwareTrojan2024}
H.~Amara, C.~Killian, D.~Chillet, and E.~Casseau, ``Mitigation of hardware trojan in noc using delta-based compression,'' in \emph{2024 IEEE 37th International System-on-Chip Conference (SOCC)}, 2024, pp. 1--6.

\bibitem{Vashist2019SecuringAW}
A.~Vashist, A.~Keats, S.~M.~P. Dinakarrao, and A.~Ganguly, ``Securing a wireless network-on-chip against jamming-based denial-of-service and eavesdropping attacks,'' \emph{IEEE Transactions on Very Large Scale Integration (VLSI) Systems}, vol.~27, pp. 2781--2791, 2019.

\bibitem{ModelingAnalysisConfluence}
S.~Bagga, R.~Gupta, and J.~Jose, ``Modelling and analysis of confluence attack by hardware trojan in noc,'' in \emph{Emerging Electronic Devices, Circuits and Systems}, C.~Giri, T.~Iizuka, H.~Rahaman, and B.~B. Bhattacharya, Eds.\hskip 1em plus 0.5em minus 0.4em\relax Singapore: Springer Nature Singapore, 2023, pp. 231--246.

\bibitem{d2021malicious}
E.~d’Afflisio, P.~Braca, and P.~Willett, ``Malicious ais spoofing and abnormal stealth deviations: A comprehensive statistical framework for maritime anomaly detection,'' \emph{IEEE Transactions on Aerospace and Electronic Systems}, vol.~57, no.~4, pp. 2093--2108, 2021.

\bibitem{elahi2024}
\BIBentryALTinterwordspacing
M.~Elahi, M.~R. Elshamy, A.-H. Badawy, M.~Fazeli, and A.~Patooghy, ``Matter: Multi-stage adaptive thermal trojan for efficiency \& resilience degradation,'' 2024. [Online]. Available: \url{https://arxiv.org/abs/2412.00226}
\BIBentrySTDinterwordspacing

\bibitem{HardwareTrojanDetectionMitigation2022}
P.~Thejaswini, G.~Vivekananda, H.~Anu, R.~Priya, K.~Prasad, and M.~Nischay, ``Hardware trojan detection and mitigation in noc using key authentication and obfuscation techniques,'' \emph{EMITTER International Journal of Engineering Technology}, pp. 370--388, 2022.

\bibitem{Tiwari2019}
B.~Tiwari, M.~Yang, Y.~Jiang, and X.~Wang, ``Effect of hardware trojan attacks on the performance of on-chip multicast routing algorithms,'' in \emph{2019 IEEE 9th Annual Computing and Communication Workshop and Conference (CCWC)}, 2019, pp. 0623--0629.

\bibitem{MitigationDenialService2016}
T.~Boraten and A.~K. Kodi, ``Mitigation of denial of service attack with hardware trojans in noc architectures,'' in \emph{2016 IEEE International Parallel and Distributed Processing Symposium (IPDPS)}, 2016, pp. 1091--1100.

\bibitem{fredricgosfand}
F.~Almario~Str{\"o}mblad and P.~Svensson, ``Securing modern integrated circuits against hardware trojans using thermal profiling : Assessing the security implications of process miniaturization,'' p.~88, 2024.

\bibitem{Tiwari2019EffectOH}
B.~Tiwari, M.~Yang, Y.~Jiang, and X.~Wang, ``Effect of hardware trojan attacks on the performance of on-chip multicast routing algorithms,'' \emph{2019 IEEE 9th Annual Computing and Communication Workshop and Conference (CCWC)}, pp. 0623--0629, 2019.

\bibitem{sachdeva2023long}
S.~Sachdeva, J.~Zhang, H.~Amrouch, and S.~X.-D. Tan, ``Long-term aging impacts on spatial on-chip power density and temperature,'' in \emph{2023 19th International Conference on Synthesis, Modeling, Analysis and Simulation Methods and Applications to Circuit Design (SMACD)}, 2023, pp. 1--4.

\bibitem{comet}
\BIBentryALTinterwordspacing
L.~Siddhu, R.~Kedia, S.~Pandey, M.~Rapp, A.~Pathania, J.~Henkel, and P.~R. Panda, ``Comet: An integrated interval thermal simulation toolchain for 2d, 2.5d, and 3d processor-memory systems,'' \emph{ACM Trans. Archit. Code Optim.}, vol.~19, no.~3, Aug. 2022. [Online]. Available: \url{https://doi.org/10.1145/3532185}
\BIBentrySTDinterwordspacing

\bibitem{7330124}
K.-C.~J. Chen, C.-H. Chao, and A.-Y.~A. Wu, ``Thermal-aware 3d network-on-chip (3d noc) designs: Routing algorithms and thermal managements,'' \emph{IEEE Circuits and Systems Magazine}, vol.~15, no.~4, pp. 45--69, 2015.

\bibitem{hotspot}
W.~Huang, S.~Ghosh, S.~Velusamy, K.~Sankaranarayanan, K.~Skadron, and M.~Stan, ``Hotspot: a compact thermal modeling methodology for early-stage vlsi design,'' \emph{IEEE Transactions on Very Large Scale Integration (VLSI) Systems}, vol.~14, no.~5, pp. 501--513, 2006.

\bibitem{zang2018markov}
D.~Zang, J.~Liu, and H.~Wang, ``Markov chain-based feature extraction for anomaly detection in time series and its industrial application,'' in \emph{2018 Chinese Control And Decision Conference (CCDC)}.\hskip 1em plus 0.5em minus 0.4em\relax IEEE, 2018, pp. 1059--1063.

\bibitem{jin2024survey}
M.~Jin, H.~Y. Koh, Q.~Wen, D.~Zambon, C.~Alippi, G.~I. Webb, I.~King, and S.~Pan, ``A survey on graph neural networks for time series: Forecasting, classification, imputation, and anomaly detection,'' \emph{IEEE Transactions on Pattern Analysis and Machine Intelligence}, 2024.

\bibitem{tang2022rethinking}
J.~Tang, J.~Li, Z.~Gao, and J.~Li, ``Rethinking graph neural networks for anomaly detection,'' in \emph{International Conference on Machine Learning}.\hskip 1em plus 0.5em minus 0.4em\relax PMLR, 2022, pp. 21\,076--21\,089.

\bibitem{5771200}
P.~Djukic and B.~Nandy, ``On threshold selection for principal component based network anomaly detection,'' in \emph{2011 Ninth Annual Communication Networks and Services Research Conference}, 2011, pp. 117--122.

\bibitem{mhatre2014temperature}
A.~D. Mhatre, \emph{Temperature evaluation of noc architectures and dynamically reconfigurable noc}.\hskip 1em plus 0.5em minus 0.4em\relax Rochester Institute of Technology, 2014.

\bibitem{pek2017confidence}
J.~Pek, A.~C. Wong, and O.~C. Wong, ``Confidence intervals for the mean of non-normal distribution: transform or not to transform,'' \emph{Open Journal of Statistics}, vol.~7, no.~3, pp. 405--421, 2017.

\bibitem{kelley2005effects}
K.~Kelley, ``The effects of nonnormal distributions on confidence intervals around the standardized mean difference: Bootstrap and parametric confidence intervals,'' \emph{Educational and Psychological Measurement}, vol.~65, no.~1, pp. 51--69, 2005.

\bibitem{esmaeili2023anomaly}
F.~Esmaeili, E.~Cassie, H.~P.~T. Nguyen, N.~O. Plank, C.~P. Unsworth, and A.~Wang, ``Anomaly detection for sensor signals utilizing deep learning autoencoder-based neural networks,'' \emph{Bioengineering}, vol.~10, no.~4, p. 405, 2023.

\bibitem{putrada2023predictive}
A.~G. Putrada, N.~Alamsyah, S.~F. Pane, M.~N. Fauzan, and D.~Perdana, ``Predictive maintenance application on machine overstrain failure with node-red and isolation forest anomaly detection,'' in \emph{2023 IEEE International Conference on Communication, Networks and Satellite (COMNETSAT)}.\hskip 1em plus 0.5em minus 0.4em\relax IEEE, 2023, pp. 64--69.

\bibitem{wang2021dl}
S.~Wang, Z.~Liu, Q.~Xiao, J.~Lin, D.~Long, and D.~An, ``Dl-rlstm: An anomaly detection framework for high dimensional time series data,'' in \emph{2021 China Automation Congress (CAC)}.\hskip 1em plus 0.5em minus 0.4em\relax IEEE, 2021, pp. 3770--3775.

\bibitem{ma2021comprehensive}
X.~Ma, J.~Wu, S.~Xue, J.~Yang, C.~Zhou, Q.~Z. Sheng, H.~Xiong, and L.~Akoglu, ``A comprehensive survey on graph anomaly detection with deep learning,'' \emph{IEEE Transactions on Knowledge and Data Engineering}, vol.~35, no.~12, pp. 12\,012--12\,038, 2021.

\bibitem{zhang2024mixpequantizationhardwarecodesign}
\BIBentryALTinterwordspacing
Y.~Zhang, M.~Wang, L.~Zou, W.~Liu, H.-L. Zhen, M.~Yuan, and B.~Yu, ``Mixpe: Quantization and hardware co-design for efficient llm inference,'' 2024. [Online]. Available: \url{https://arxiv.org/abs/2411.16158}
\BIBentrySTDinterwordspacing

\bibitem{sniper}
\BIBentryALTinterwordspacing
T.~E. Carlson, W.~Heirman, S.~Eyerman, I.~Hur, and L.~Eeckhout, ``An evaluation of high-level mechanistic core models,'' \emph{ACM Trans. Archit. Code Optim.}, vol.~11, no.~3, Aug. 2014. [Online]. Available: \url{https://doi.org/10.1145/2629677}
\BIBentrySTDinterwordspacing

\bibitem{parsec}
\BIBentryALTinterwordspacing
C.~Bienia, S.~Kumar, J.~P. Singh, and K.~Li, ``The parsec benchmark suite: characterization and architectural implications,'' in \emph{Proceedings of the 17th International Conference on Parallel Architectures and Compilation Techniques}, ser. PACT '08.\hskip 1em plus 0.5em minus 0.4em\relax New York, NY, USA: Association for Computing Machinery, 2008, p. 72–81. [Online]. Available: \url{https://doi.org/10.1145/1454115.1454128}
\BIBentrySTDinterwordspacing

\end{thebibliography}
